\documentclass[aps,prl,showpacs,twocolumn]{revtex4-1}
\usepackage{bm,color,amsmath,amssymb,mathrsfs,latexsym,graphicx,psfrag}

\begin{document}
\title{Magnetic order and paramagnetic phases in the quantum
  $J_1$-$J_2$-$J_3$ honeycomb model} 

\author{Johannes Reuther${}^1$}
\author{Dmitry A. Abanin${}^{2,3}$}
\author{Ronny Thomale${}^{3,4}$}

\affiliation{${}^1$Institut~f\"ur~Theorie~der~Kondensierten~Materie,~Karlsruhe Institute of Technology,~76128 Karlsruhe,~Germany} 
\affiliation{${}^2$Princeton Center for Theoretical Science, Princeton University, Princeton, NJ 08544}
\affiliation{${}^3$Department of Physics, Princeton University, Princeton, NJ 08544, USA} 
\affiliation{${}^4$Microsoft Station Q, Elings Hall, University of California at Santa Barbara, Santa Barbara CA 93106, USA}


\begin{abstract}
%
Recent work shows that a quantum spin liquid can arise in realistic fermionic models on a honeycomb lattice. We study the quantum spin-1/2 Heisenberg honeycomb model, considering couplings $J_1$, $J_2$, and $J_3$ up to third nearest neighbors. We use an unbiased pseudofermion functional renormalization group method to compute the magnetic susceptibility and determine the ordered and disordered states of the model. Aside from antiferromagnetic, collinear, and spiral order domains, we find a large paramagnetic region at intermediate $J_2$ coupling. For larger $J_2$ within this domain, we find a strong tendency to staggered dimer ordering, while the remaining paramagnetic regime for low $J_2$ shows only weak plaquet and staggered dimer response. We suggest this regime to be a promising region to look for quantum spin liquid states when charge fluctuations would be included.
\end{abstract}
\date{\today}


\maketitle

The search for a quantum spin liquid phase in nature ever since has been a complicated task not only from experiment, but also from theory~\cite{anderson87s1197,balents10n199}. This stems from the fact that the properties of a quantum spin liquid are very peculiar: quantum fluctuations have to form a many-body singlet state without long-range correlations of any kind of operator. Since the notion of frustration in quantum spin systems which is the main resource of fluctuations to accomplish a magnetically disordered quantum state at zero temperature, studies of a plethora of spin Hamiltonians on different lattices tell us that even most frustrated quantum systems tend to establish some sort of long-range correlations such as seen in a valence bond solid phase: while local spin operator correlations rapidly fall off there, long-range dimer-dimer correlations spoil the phenomenology of a quantum spin liquid.

The remarkable numerical studies by Meng et al.~\cite{meng-10n847} are a fortunate exception. In their work, they report on the first unambiguous discovery of a genuine spin liquid phase from a generic microscopic model. They consider the Hubbard model on the honeycomb lattice by Monte Carlo methods and find a spin-gapped phase at $U/t=4.3$ which shows no long-range correlations of any kind, neither charge density wave, superconductivity or even spin solid-type correlations such as that of a valence-bond crystal formation. Moreover, the study finds a clear excitation gap and no symmetry breaking of any lattice symmetry or parity $P$ and time-reversal $T$, which already excludes chiral spin liquids and algebraic spin liquids.

One path of further understanding the magnetic quantum phases on the honeycomb lattice is the development of effective descriptions for the Hubbard model itself such as gauge theory~\cite{hermele07prb035125} and slave boson mean field theory methods~\cite{wang10prb024419, vaezi-cm}. 
Another direction, however, which we pursue in this Letter is to analyze the Gutzwiller-projected Hubbard model on the honeycomb lattice. Specifically, we consider Heisenberg spin couplings up to third nearest neighbors labeled as the $J_1$-$J_2$-$J_3$ model. The motivation for this is two-fold. First, the $J_1$-$J_2$-$J_3$ model projects onto the square root fraction of the Hilbert space of single site
occupancy where only spin modes are present, which enables us to analyze the model through methods designed for this setup. Second, in reverse, a detailed understanding of the $J_1$-$J_2$-$J_3$ phase diagram will eventually help to identify which aspects of possible quantum phases may be explained through spin fluctuations only and which may necessitate the effect of charge fluctuations.

\begin{figure}[t]
\begin{minipage}{0.99\linewidth}
\includegraphics[width=\linewidth]{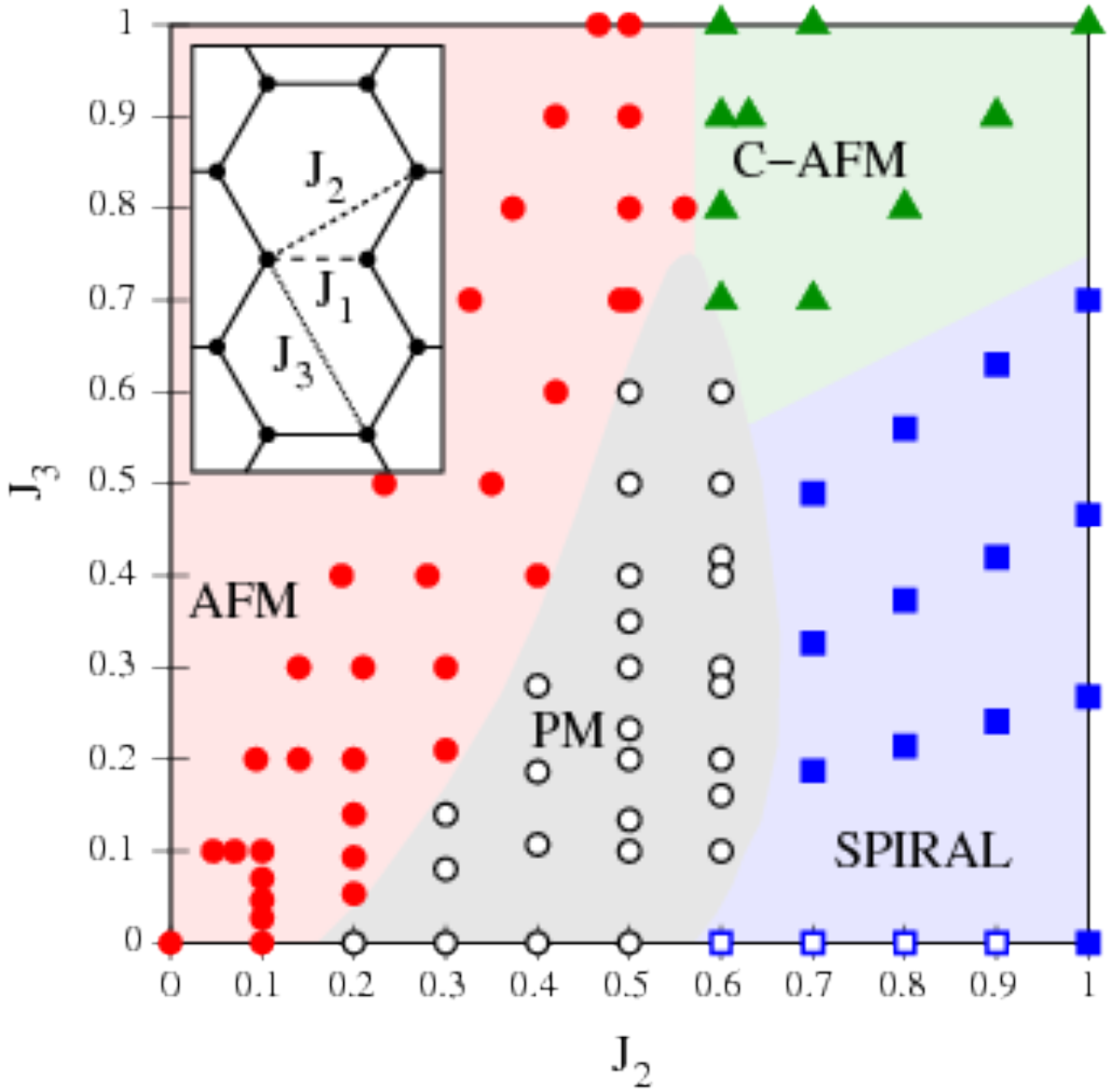}
\end{minipage}
\caption{(Color online) Phase diagram of the $J_1$-$J_2$-$J_3$ honeycomb model via PFFRG (with exchange couplings depicted in the inset). The points represent parameter settings which we have computed. In the depicted $J_2$-$J_3$ range we find AFM N\'eel order (red circles), collinear order (C-AFM, green triangles), spiral order (blue squares) and a paramagnetic phase (open circles). The spiral order phase partly shows incommensurability shifts (open blue squares) from the spiral phase for dominant $J_2$ (see also Fig.~\ref{pic2}).}
\label{pic1}
\end{figure}

We consider the Hamiltonian
\begin{equation}
H_{\text{HCM}}=J_1 \sum_{\langle i, j \rangle}  \vec{S}_i \vec{S}_j +
J_2 \hspace{-5pt} \sum_{\langle
  \langle i,j \rangle \rangle} \vec{S}_i \vec{S}_j+ J_3 \hspace{-8pt} \sum_{\langle \langle \langle i, j
\rangle \rangle \rangle} \vec{S}_i \vec{S}_j,
\label{eq:model}
\end{equation}
where the sums extend over nearest, second nearest and third nearest neighbors, respectively (see Fig.~\ref{pic1}). This model is expected to describe a class of magnetic materials with a honeycomb lattice, one example being $\beta-\text{Cu}_2\text{V}_2\text{O}_7$~\cite{tsirlin-10prb144416}. $J_2$ and $J_3$ are given in units of $J_1$. 
The solution of the classical $J_1$-$J_2$-$J_3$ model has been known for a long time~\cite{rastelli-79pb1,katsura-86jsm381}. For small $J_2$ the system is N\'eel-ordered, which is commensurate with the honeycomb lattice and preserves the sublattice $120^{\circ}$ degrees rotational symmetry. For sufficiently low $J_3$ and beyond a threshold of $J_2$, the system resides in a spiral phase. For high $J_2$ and $J_3$, it is energetically preferred to order in a collinear phase where spins along zigzag chains align ferromagnetically while neighboring zigzag chains exhibit antiparallel spin orientation (there are three degenerate collinear configurations).
 
In order to obtain an adequate quantum phase diagram of the model defined in~(\ref{eq:model}), there are not many suitable methods available, some of which predominantly focused on the $J_1$-$J_2$ line. Exact diagonalization (ED) studies are very helpful, as arbitrary $m$-point correlation functions can be computed in principle. However, except of the valence bond crystal domain where the dimer Hilbert space projection~\cite{poilblanc-09cm0724} is justified~\cite{mosadeq-cm}, ED cannot reach sufficient system sizes to adequately determine all phase regimes~\cite{fouet-01epj241,cabra-cm,mosadeq-cm}. Linear spin wave theory gives a qualitative tendency where the quantum corrections lead to, but is still strongly biased towards the classical limit~\cite{fouet-01epj241,cabra-cm,mulder-10prb214419}. Similar shortcomings apply to Schwinger boson mean field theory~\cite{mattsson-94prb3997, cabra-cm}. A promising direction has recently been given by variational Monte Carlo (VMC) schemes, where energies of the antiferromagnetic (AFM) ordered trial state as well as several spin liquid (solid) candidates have been compared for $J_3=0$, $0<J_2<0.5$~\cite{clark-cm}. There, a transition point of $J_2=0.08$ is found from the AFM ordered phase to the spin liquid phase which breaks no rotational invariance. 
It is followed by a transition to a dimerized spin solid phase at $J_2=0.3$ which breaks rotational invariance of the lattice. 
It is hence likely that the disordered phase itself, already for the $J_1$-$J_2$ line and as such definitely for the full phase diagram with finite $J_3$, contains different types of paramagnetic phases, which we will investigate in the following.

\begin{figure*}[t]
\begin{minipage}{0.99\linewidth}
\includegraphics[width=0.85\linewidth]{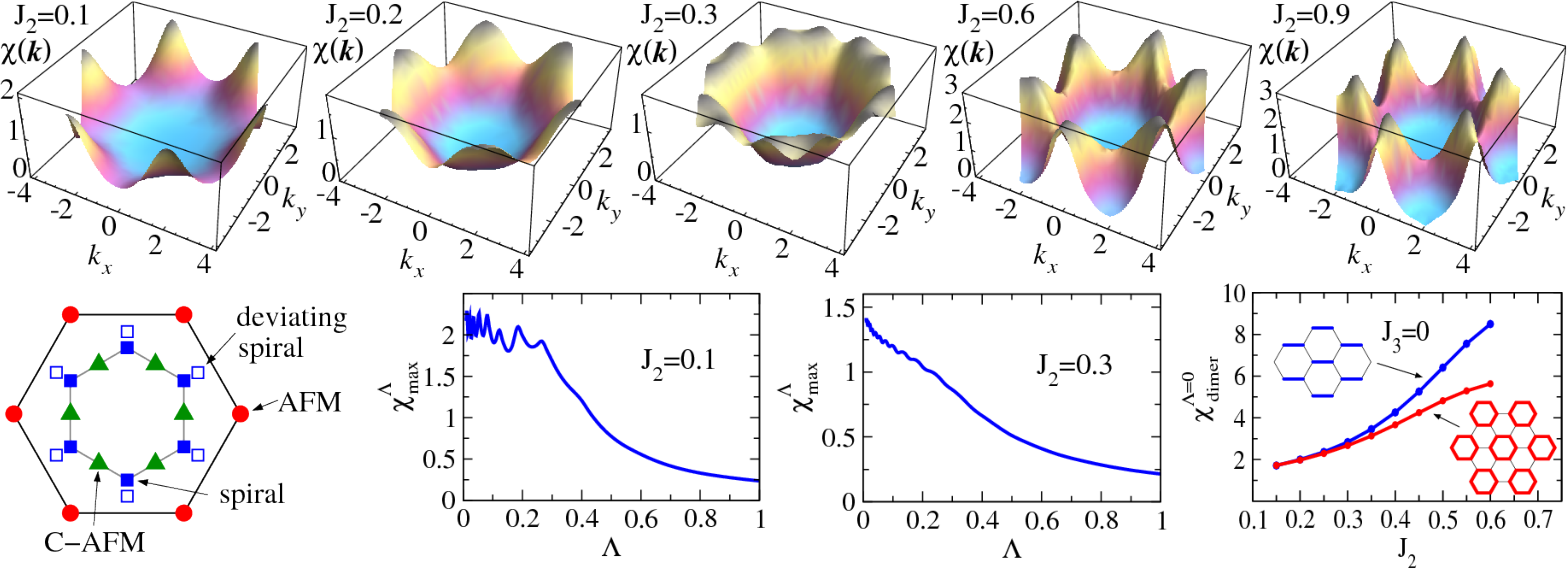}
\end{minipage}
\caption{(Color online) The Heisenberg honeycomb model along the $J_3=0$ axis. Upper line: $J_2$-sweep of the static $\mathbf{k}$-space resolved susceptibility. In magnetic phases ($J_3=0.1$) the susceptibility is depicted at a cutoff scale just before the instability breakdown, otherwise the correlations are derived from the final flow values at $\Lambda=0$. Susceptibilities are always given in units of $\frac{1}{J_1}$. Lower line: Left: Wave-vector positions for different types of magnetic order in the extended BZ. The inner hexagon marks the first BZ. Middle: Examples for the flow behavior of the leading susceptibility component. $J_2=0.1$ shows a representative example of an RG flow displaying order and $J_2=0.3$ for a disordered phase. Right: Staggered and plaquet dimer response along the $J_3=0$ axis. The insets show the staggered and plaquet dimerization patterns.}
\label{pic2}
\end{figure*}

We employ the pseudo-fermion functional renormalization group (PFFRG)~\cite{reuther-10prb144410,reuther-10prb024402} to identify magnetically ordered and paramagnetic phases and compute the groundstate magnetic susceptibility to determine the magnetic order parameter in the ordered phases as well as the dominant fluctuation profile in the paramagnetic phases. 
Our starting point is the pseudofermion representation of spin-1/2 operators $S^{\mu} = 1/2 \sum_{\alpha\beta} f_{\alpha}^{\dagger} \sigma_{\alpha\beta}^{\mu} f_{\beta}$, ($\alpha,\beta = \uparrow,\downarrow$, $\mu = x,y,z$) with the fermionic operators $f_{\uparrow}$ and $f_{\downarrow}$ and the Pauli-matrices $\sigma^{\mu}$. This representation enables us to apply Wick's theorem leading to standard Feynman many-body techniques. 
Quantum spin models are inherently strongly coupled models, requiring an infinite self-consistent resummation theory. In this context
FRG~\cite{wetterich93plb90,morris-94ijmpa2411,honerkamp-01prb035109,hedden-04jpcm5279,salmhofer-04ptp943,reuther-10prb024402,reuther-11prb064416} provides a systematic summation in different interaction channels by generating equations for the evolution of all one-particle irreducible $m$-particle vertex functions under the flow of an IR frequency cutoff $\Lambda$. In order to reduce the infinite hierarchy of equations to a closed set, we restrict the computation to the full set of one-loop parquet diagrams and their vertex corrections up to infinite order, and additionally include certain two-loop contributions that are essential to induce a self-consistent resummation procedure exceeding the perturbative limit~\cite{katanin04prb115109,reuther-10prb024402}. The parquet diagrams include graphs that favor magnetic order and those that favor disorder tendencies such that in total the method provides an adequate treatment of order and disorder fluctuations~\cite{anderson52pr694,marston-89prb11538,reuther-10prb024402}. From the two-particle scattering vertex we obtain the spin-spin correlation function (or spin susceptibility), which is the central outcome of our PFFRG. A magnetic ordering instability is initially signalled by a strong rise of the susceptibility associated with this order at some finite scale of $\Lambda$. The onset of spontaneous long-range order is signalled by a sudden breakdown of the smooth flow (as shown for $J_2=0.1$ in Fig.~\ref{pic2}). In the momentum-resolved magnetic susceptibility $\chi(\mathbf{k})$, where $\mathbf{k}$ is defined along $k_{x,y}$ components in the extended Brillouin zone (BZ), the different magnetic ordering patterns manifest as peaks depicted in Fig.~\ref{pic2}. Note that due to the two-atomic unit cell, the spin susceptibility does not have the periodicity of the first BZ but rather of the extended (second) BZ. Adding small dimer-field perturbations to the Hamiltonian, we are able to calculate dimer responses for a given dimer configuration (Fig.~\ref{pic2}).
For the present study our PFFRG algorithm internally deals with spin-spin correlations up to a length of $9$ lattice constants corresponding to a correlation area of $181$ sites which provides an adequate description of the model. 

Fig.~\ref{pic1} shows the quantum phase diagram of the $J_1$-$J_2$-$J_3$ model~(\ref{eq:model}). For dominant $J_1$ the system displays AFM order which persists longer against $J_2$ for finite $J_3$ as $J_3$ cooperates with $J_1$. Increasing $J_2$ (for not too large $J_3$) we observe a melting of the order and the appearance of a rather large paramagnetic region. Above $J_2\approx0.6$ the system is characterized by presumably weak magnetic order and very small ordering instability scales which are hard to resolve numerically. However, as we enter this region by increasing $J_2$, we observe the appearance of clean magnetic response peaks which we interpret as the onset of weak magnetic order. From the peak positions in $\mathbf{k}$-space we divide this region into a collinear ordered phase for large $J_3$ and a spiral ordered phase for smaller $J_3$ (Fig.~\ref{pic1}). 
Throughout parameter space, the spiral phase is very close to $120^{\circ}$-N\'eel order on both honeycomb sublattices except for a small region near the $J_2$ axis where the wave vector deviates from commensurability (Fig.~\ref{pic2}). From the perspective of the degenerate classical spiral phase, this corresponds to a pinning of the order due to quantum fluctuations.
%
A pronounced jump of the leading susceptibility wave-vector is seen as we cross the transition between AFM and C-AFM order, pointing to a first order transition. The observations are consistent with the classical result except for the fact that the transition between AFM and C-AFM order is shifted towards higher $J_{2,\textnormal{c}}\approx0.57$ compared to $J_{2,\textnormal{c}}^{\text{classical}}=0.5$ due to quantum corrections included in our calculation.

We now focus on selected cuts through parameter space. We first investigate how the fluctuation profile changes along the $J_3=0$ line (Fig.~\ref{pic2}). For small $J_2$ AFM order manifests itself in peaks at the corners of the extended BZ  and a characteristic instability breakdown of the flow (lower line of Fig.~\ref{pic2}). As we increase $J_2$, the AFM peaks rapidly decrease and from the disappearance of unstable flow behavior we estimate the transition to be at $J_2\approx0.15$. Inside the paramagnetic phase, such as at $J_2=0.3$ depicted in Fig.~\ref{pic2}, no clear peak structure is visible and the susceptibility flow remains stable up to $\Lambda\rightarrow0$. Around $J_2\approx0.6$ spiral order peaks emerge at wave vectors slightly shifted from the commensurate positions towards larger $|\mathbf{k}|$. This feature is consistent with a large-$S$ expansion~\cite{mulder-10prb214419} which allows to select specific wave vectors out of a classical manifold of degenerate momenta. Upon further increasing $J_2$, the peak positions approach commensurability, i.e., the sublattices effectively decouple and individually exhibit $120^{\circ}$-N\'eel order. During the flow, these susceptibility peaks emerge at very small $\Lambda$-scales, giving us indication of a very weak magnetization. While the transition between paramagnetism and spiral order is a bit smeared out at small $J_3$, the onset of response peaks occurs more abruptly at larger $J_3$. 

\begin{figure*}[t]
\begin{minipage}{0.99\linewidth}
\includegraphics[width=0.85\linewidth]{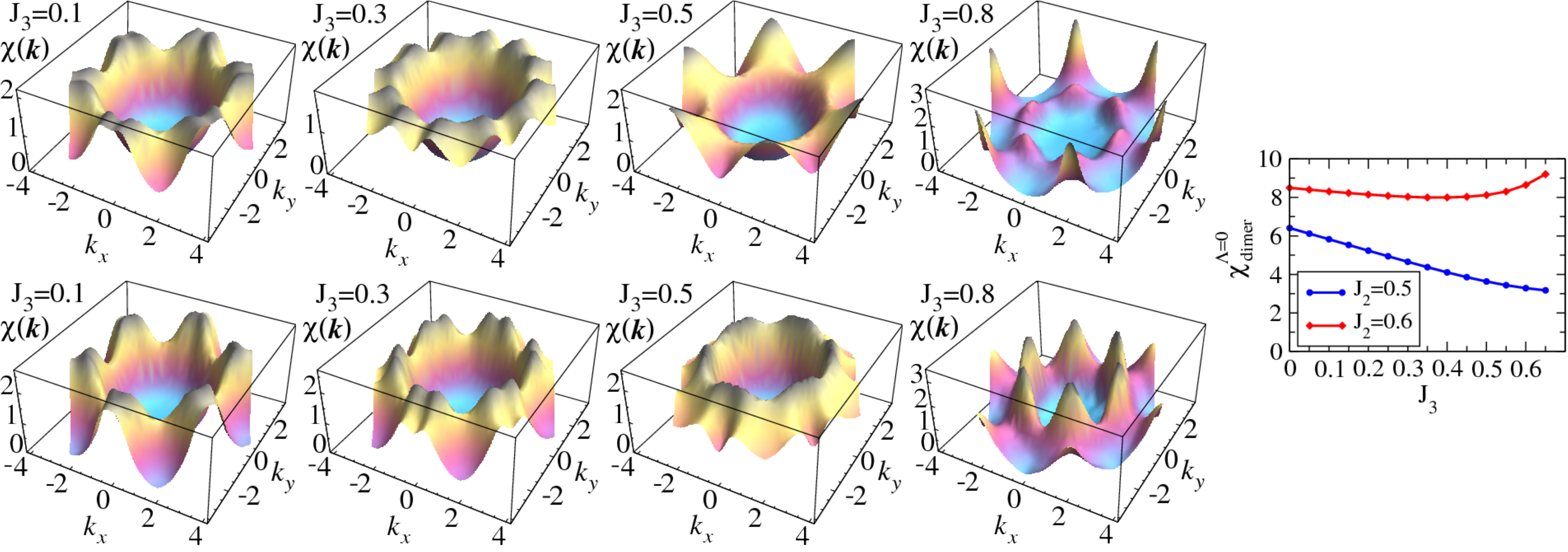}
\end{minipage}
\caption{(Color online) $J_3$ sweep of the susceptibility for $J_2=0.5$ (upper line) and $J_2=0.6$ (lower line). As $J_3$ is increased, the disordered phase enters the AFM phase for $J_2=0.5$ and enters the C-AFM phase for $J_2=0.6$. Right: Corresponding staggered dimer responses (plaquet dimer responses are much weaker and not shown). We find an increased propensity to staggered dimer order for larger $J_2$.}
\label{pic3}
\end{figure*}

To resolve more information about the correlations in the disordered phase, we compute the staggered and plaquet dimer response, i.e., the (dimensionless) factor of amplification of an external dimer-field perturbation exerted on the system. In doing so we can distinguish between parameter regimes of different dimer fluctuation strength. 
As we sweep through the paramagnetic phase at $J_3=0$ we find that the staggered dimer response is dominant for higher $J_2$ while plaquet and staggered dimer response compete for lower $J_2$ (Fig.~\ref{pic2}). The absolute dimer response amplitudes are smaller for lower $J_2$. This is in qualitative agreement with VMC~\cite{clark-cm} as well as with ED studies~\cite{mosadeq-cm}. It supports the view that if at all the system may form a spin liquid phase around this domain which is also the parameter regime related to the honeycomb Hubbard model from a strong coupling expansion. There, charge fluctuations which are neglected in~\eqref{eq:model} may be sufficient to destroy the comparably low dimer ordering tendency.

In addition, we investigate parameter lines varying $J_3$ for intermediate $J_2$ through the disordered regime. The fluctuation profiles and dimer responses for the paramagnetic regime at finite $J_3$ are shown in Fig.~\ref{pic3}. We see that at small $J_3$ the fluctuation profiles for $J_2=0.5$ and $J_2=0.6$ are very similar but differ more with increasing $J_3$ until eventually the $J_2=0.5$ line leads into the AFM ordered phase while collinear order emerges on the $J_2=0.6$ line. The paramagnetic phase shows rather complicated susceptibility profiles which have lost most signature of ordering fluctuations. A typical feature is the ringlike shape as seen e.g. at $(J_2, J_3)=(0.6,0.5)$. An intuitive reason for a quantum disordered phase is already indicated from the classical limit where the point $(J_2, J_3)=(0.5,0.5)$ is tricritical with three competing ordering tendencies. From the dimer responses along the $J_2=0.5$ and $J_2=0.6$ line we find that rather independent of $J_3$, staggered dimer ordering tendency is more efficiently established by larger $J_2$. 



{\it Note added.} When this manuscript was completed we became aware of an independent work providing an analysis of model~\eqref{eq:model} through joint mean field and exact diagonalization techniques~\cite{albuquerque-cm1102}. Several similar findings show a good correspondence of both approaches. 

\begin{acknowledgments}  
We thank S.~Capponi, B.~Clarke, A.~Hackl, M.~Hastings, A.~L\"auchli, C.~Lhuillier, A.~Muramatsu, K.~P. Schmidt, R.~Singh, S.~Trebst, S.~Wessel, and, in particular, P.~W\"olfle for discussions. RT thanks the KITP workshop 'Disentangling Quantum Many-body Systems: Computational and Conceptual Approaches' for hospitality. JR is supported by DFG-FOR 960. RT is supported by DFG-SPP 1458/1 and the Humboldt Foundation.
\end{acknowledgments}


\begin{thebibliography}{10}

\bibitem{anderson87s1197}
P.~W. Anderson, Science {\bf 235},  1196  (1987).

\bibitem{balents10n199}
L. Balents, Nature {\bf 464},  199  (2010).

\bibitem{meng-10n847}
Z.~Y. Meng, T.~C. Lang, S. Wessel, F. Assaad, and A. Muramatsu, Nature {\bf
  464},  847  (2010).

\bibitem{hermele07prb035125}
M. Hermele, Phys. Rev. B {\bf 76},  035125  (2007).

\bibitem{wang10prb024419}
F. Wang, Phys. Rev. B {\bf 82},  024419  (2010).

\bibitem{vaezi-cm}
A. Vaezi and X.~G. Wen, arXiv:1010.5744.

\bibitem{tsirlin-10prb144416}
A.~A. Tsirlin, O. Janson, and H. Rosner, Phys. Rev. B {\bf 82},  144416
  (2010).

\bibitem{rastelli-79pb1}
E. Rastelli, A. Tassi, and L. Reatto, Physica~B {\bf 97},  1  (1979).

\bibitem{katsura-86jsm381}
S. Katsura, T. Ide, and T. Morita, J. Stat. Phys. {\bf 42},  381  (1986).

\bibitem{poilblanc-09cm0724}
D. Poilblanc, M. Mambrini, and D. Schwandt, arXiv:0912.0724.

\bibitem{mosadeq-cm}
H. Mosadeq, F. Shahbazi, and S.~A. Jafari, arXiv:1007.0127.

\bibitem{fouet-01epj241}
J.~B. Fouet, P. Sindzingre, and C. Lhuillier, Eur.~Phys.~J.~B {\bf 20},  241
  (2001).

\bibitem{cabra-cm}
D.~C. Cabra, C.~A. Lamas, and H.~D. Rosales, arXiv:1003.3226.

\bibitem{mulder-10prb214419}
A. Mulder, R. Ganesh, L. Capriotti, and A. Paramekanti, Phys. Rev. B {\bf 81},
  214419  (2010).

\bibitem{mattsson-94prb3997}
A. Mattson, P. Fr\"ojdh, and T. Einarsson, Phys. Rev. B {\bf 49},  3997
  (1994).

\bibitem{clark-cm}
B.~K. Clark, D.~A. Abanin, and S.~L. Sondhi, arXiv:1010.3011.

\bibitem{reuther-10prb144410}
J. Reuther and P. W\"olfle, Phys. Rev. B {\bf 81},  144410  (2010).

\bibitem{reuther-10prb024402}
J. Reuther and R. Thomale, Phys. Rev. B {\bf 83},  024402  (2011).

\bibitem{wetterich93plb90}
C. Wetterich, Phys. Lett. B {\bf 301},  90  (1993).

\bibitem{morris-94ijmpa2411}
T.~R. Morris, Int. J. Mod. Phys. A {\bf 9},  2411  (1994).

\bibitem{honerkamp-01prb035109}
C. Honerkamp, M. Salmhofer, N. Furukawa, and T.~M. Rice, Phys. Rev. B {\bf 63},
   035109  (2001).

\bibitem{hedden-04jpcm5279}
R. Hedden, V. Meden, T. Pruschke, and K. Sch\"onhammer, J. Phys.: Condes.
  Matter {\bf 16},  5279  (2004).

\bibitem{salmhofer-04ptp943}
M. Salmhofer, C. Honerkamp, W. Metzner, and O. Lauscher, Prog. Theor. Phys.
  {\bf 112},  943  (2004).

\bibitem{reuther-11prb064416}
J. Reuther, P. W\"olfle, R. Darradi, W. Brenig, M. Arlego, and J. Richter,
  Phys. Rev. B {\bf 83},  064416  (2011).

\bibitem{katanin04prb115109}
A.~A. Katanin, Phys. Rev. B {\bf 70},  115109  (2004).

\bibitem{anderson52pr694}
P.~W. Anderson, Phys. Rev. {\bf 86},  694  (1952).

\bibitem{marston-89prb11538}
J.~B. Marston and I. Affleck, Phys. Rev. B {\bf 39},  11538  (1989).

\bibitem{albuquerque-cm1102}
A.~F. Albuquerque, D. Schwandt, B. Hetenyi, S. Capponi, M. Mambrini, and A.~M.
  L\"auchli, arXiv:1102.5325.

\end{thebibliography}

\end{document}